\documentclass[10pt,journal,compsoc]{IEEEtran}

\ifCLASSOPTIONcompsoc
  \usepackage[nocompress]{cite}
\else
  \usepackage{cite}
\fi

\usepackage[pdftex]{graphicx}
\usepackage[caption=false,font=footnotesize,labelfont=sf,textfont=sf]{subfig}
\usepackage{svg}
\usepackage{amsmath}
\usepackage{algorithmic}
\usepackage{url}
\usepackage{xcolor}
\usepackage[normalem]{ulem}
\usepackage{arydshln}
\usepackage[caption=false, font=footnotesize]{subfig}

\begin{document}

%
\title{Conditioned Source Separation for Musical Instrument Performances}
%
%
%
%

\author{Olga~Slizovskaia, 
        Gloria~Haro, 
        and~Emilia~G\'omez
\IEEEcompsocitemizethanks{\IEEEcompsocthanksitem O.Slizovskaia and G.Haro are with the Department of Information and Communication Technologies, Universitat Pompeu Fabra, Barcelona,
Spain, 08018.\protect\\
E-mail: olga.slizovskaia@upf.edu, gloria.haro@upf.edu
\IEEEcompsocthanksitem E.G\'omez is with the Joint Research Centre, European Commission, Seville, Spain, 41092; and the Department of Information and Communication Technologies, Universitat Pompeu Fabra, Barcelona,
Spain, 08018.}
\thanks{Manuscript received March 23, 2020; revised October 13, 2020; accepted April 4, 2021.}}

%
%

\markboth{IEEE/ACM TRANSACTIONS ON AUDIO, SPEECH, AND LANGUAGE PROCESSING, VOL. XX, 2021}%
{Shell \MakeLowercase{\textit{et al.}}: Bare Demo of IEEEtran.cls for Computer Society Journals}
%



\IEEEtitleabstractindextext{%
\begin{abstract}
In music source separation, the number of sources may vary for each piece and some of the sources may belong to the same family of instruments, thus sharing timbral characteristics and making the sources more correlated. This leads to additional challenges  in the source separation problem.
This paper proposes a source separation method for multiple musical instruments sounding simultaneously and explores how much additional information apart from the audio stream can lift the quality of source separation. We explore conditioning techniques at different levels of a primary source separation network and utilize two extra modalities of data, namely presence or absence of instruments in the mixture, and the corresponding video stream data.
\end{abstract}

\begin{IEEEkeywords}
Single Channel Source Separation, Audio-Visual Analysis, Conditioned Neural Networks
\end{IEEEkeywords}}

\maketitle

\IEEEdisplaynontitleabstractindextext
\IEEEpeerreviewmaketitle

\IEEEraisesectionheading{\section{Introduction}\label{sec:introduction}}

\IEEEPARstart{H}umans combine sensory information of different origins to obtain a comprehensive picture of the world. Actually, the information that we perceive with different senses often originates from the same object or event at the physical level. Thus, sound is a vibration which is commonly produced by an object's movement. In the case of musical instruments sounds, we even can see the movements and associate a particular sound with its source \cite{li2017see, zhao2019soundofmotions}. Also, each of those instruments has its own unique visual characteristics such as shape and color which help us recognize them. While experiencing online listening through video streaming, we are often given all this information together with prior knowledge we may have about a piece, instrumentation or an artist. All this helps us with the interpretation of the music we are exposed to, such that, while listening, we can focus on the individual sources of the sound and identify them.
Moreover, such phenomenon as synesthesia can also support audio-visual correspondence studies. Thus, some articles report correlations between loudness and visual size/light intensity, as well as musical timbre and arbitrary visual shapes \cite{adeli2014audiovisual}. 

In this paper, we focus on Single Channel Source Separation (SCSS). This task is usually solved in the audio domain, but in this work we explore the effects of integrating two additional kinds of context data, namely instruments labels and their visual properties.

We work with audio-visual recordings of musical ensembles with several families of instruments that can be commonly found in a symphonic orchestra such as strings, woodwinds and brass instruments, that is to say, mostly chamber music. Source separation with such a setup is known to be a very challenging task, and has been approached in the past with multi-channel score-informed methods \cite{miron2016score} and timbre-informed methods \cite{carabias2013nonnegative}. It is worth noticing that previous studies operated on multi-channel recordings with different instruments present in each channel, as there are no large datasets with ground truth audio available for each target source.

The music source separation problem has several challenges, to mention a few:  
\begin{itemize}
    \item The instruments within a family can be very similar to one another;
    \item The number of sources in the mixture is unknown in advance;
    \item There is a considerable amount of overlap in time and frequency between sources;
    \item Even for instruments which have essentially different timbre, tone color, and different performance techniques, such as clarinet and viola, some musicians may mimic a sound of one while playing another \cite{lee2004analysis}.
\end{itemize}

Different recent works have tackled the music source separation problem using audio-visual data and deep learning techniques \cite{zhao2018soundofpixels, zhao2019soundofmotions, gao2018learning, gao2019co, xu2019recursive}. Due to the lack of ground-truth data, all of these works have followed the \textit{mix-and-separate} strategy for training, proposed in \cite{zhao2018soundofpixels}, which consists of artificially creating mixes by adding the sound of individual sources. Similarly, the same kind of unrealistic mixtures is used for the quantitative evaluation. In addition, previous studies only considered mixes of only two sources, where instruments from the same family were rarely present. These restrictions affect the generality of the proposed approaches. In fact, separating different musical instruments playing the same piece, i.e. a real mixture, is a more challenging task than these unrealistic scenarios, since the different audio sources are synchronized and playing in harmony.

In this work, we also adopt the mix-and-separate technique while training; again, because of the absence of appropriate datasets for our purposes. But contrarily to previous works, we evaluate and compare the different methods in a more realistic scenario, thanks to the URMP dataset \cite{urmpdataset}.

As for combining different modalities of information, the key technical problem for many years has been the large gap (both in dimensions and content) between representations of the auditory and visual modalities \cite{kidron2005pixels}. One of the common approaches consisted in feature construction followed by dimensionality reduction \cite{kidron2005pixels, hershey2000audio}. Deep learning techniques and their powerful representation capability have helped in reducing the heterogeneity and dimensionality gap among different modalities \cite{guo2019deep}. However, a proper way of fusing different data representations still remains an issue \cite{baltruvsaitis2018multimodal,ramachandram2017deep,michelsanti2020overview}.

In this regard, this work explores conditioning techniques at different levels of a primary source separation network. We are not the first ones to propose Conditioned-U-Net for source separation or audio-visual source separation \cite{gao2019co, zhao2018soundofpixels, zhao2019soundofmotions, korbar2018cooperative, meseguer2019conditioned}. However, unlike prior approaches, we conduct a thorough study identifying the optimal type of conditioning and comparing possible conditioning strategies with two types of context data: the presence or absence of instruments in the mixture and the video stream data. We train on mixtures of up to 7 sources, and evaluation is carried out on real-world mixtures from the URMP\cite{urmpdataset} dataset which has up to 4 different instruments per piece, often from the same family.
We should note that we only address the separation of sources of different type (in our case different musical instruments) and thus, two or more instances of the same instrument are treated as a single source. 

The complexity of the task allows for the present approach to be used as a baseline for future research. 
In order to facilitate that, the present study is reproducible as we provide pretrained models, code, data and all the training parameters. The supplementary materials and examples are available at \url{https://veleslavia.github.io/conditioned-u-net/}. 
%

This paper provides an overview of existing techniques for source separation, audio-visual methods and conditioning strategies in Section \ref{ss:sota}, indicating relations and differences with the present work. In Section \ref{ss:approach} we formalize our approach for performing source separation conditioned on context data. This is followed by Section \ref{ss:experiments} where we describe the experimental setup and implementation details.
Finally, we discuss the obtained results in Section \ref{ss:results} and provide conclusions in Section \ref{ss:conclusion}.


\section{Related work}\label{ss:sota}

\subsection{Single channel source separation}

Single channel source separation (SCSS) consists of estimating the individual sources $x_i$ given a mono time-domain mixture signal $y$ of $N$ sources:

\begin{equation}
y(t) = \sum_{i=1}^{N} x_i(t).
\end{equation}

Instead of predicting time-domain signals, a general approach for solving SCSS involves the estimation of $N$ masks for Short-Term Fourier transform (STFT) values of the mixture. In this case, we consider a time-frequency representation of the mixture $ \boldsymbol{Y} $ and of the sources $ \boldsymbol{X_i} $, and the goal of the source separation method is to learn a real-valued (or complex-valued) mask $M_i$ for each source $i$ .

Let us denote the STFT magnitude of $\boldsymbol{X_i}(\tau, \omega)$ and $\boldsymbol{Y}(\tau, \omega)$  by $| \boldsymbol{X_i} |$ and $| \boldsymbol{Y} |$, respectively, with frequency $\omega$ and time frame $\tau$.
In this work, we only consider two types of \textit{real-valued} masks, namely \textit{ideal ratio or soft} masks $M_i^{ir}$:
\begin{equation}
    M_i^{ir}(\tau, \omega) = \frac{|\boldsymbol{X_i}(\tau, \omega)|}{|\boldsymbol{Y}(\tau, \omega)|},
\end{equation}
and \textit{ideal binary} masks\footnote{\url{https://github.com/sigsep/sigsep-mus-oracle/blob/master/IBM.py}}
$M_i^{ib}$:
\begin{equation}
\begin{gathered}
M_i^{ib}(\tau, \omega) =
    \begin{cases}
      1, & \text{if}\  \frac{|\boldsymbol{X_i}(\tau, \omega)|}{|\boldsymbol{Y}(\tau, \omega)| - |\boldsymbol{X_i}(\tau, \omega)|} \geq 1 \\
      0, & \text{otherwise},
    \end{cases}
\end{gathered}
\end{equation}

We obtain the STFT magnitude values of separated sources by multiplying the STFT magnitude of the mixture by the estimated masks $\hat{M}_i$, i.e., $|\boldsymbol{\hat{X}_i}| = \hat{M}_i \odot |\boldsymbol{Y}|$, where $\odot$ denotes the Hadamard product (element-wise multiplication). Then, the waveforms of the source signals are recovered by applying the inverse STFT transform on the predicted magnitude $|\boldsymbol{\hat{X}_i}|$ and using the phase of the mixture $\boldsymbol{Y}$.

The mask estimation step has always been an essential component of model-based source separation algorithms \cite{carabias2013nonnegative, miron2016score, parekh2018guiding, carabias2011musical, ozerov2009multichannel, virtanen2007monaural}. Consecutively, the masking-based approach for training neural networks has received a lot of attention recently and has been very successful in SCSS \cite{pritish2017deepconv, wisdom2019differentiable, jansson2017singing}. Additional constraints such as mixture-consistency and STFT consistency on the estimated sources have proven to improve mask-based SCSS networks \cite{wisdom2019differentiable}. Despite the fact that most of the existing work has focused on estimating binary or ratio masks, the direct estimation of STFT magnitude values has also been used in practice \cite{doire2019interleaved} together with loss function computation in time-frequency \cite{stoter19} or time domain \cite{kavalerov2019universal} while internally estimating the masks.

It's worth noting that the set of methods which has been successfully used in source separation is very diverse, and the optimal choice of an architecture remains an open research question. Some examples include LSTMs~\cite{luo2018tasnet} and BLSTMs \cite{uhlich2017improving, stoter19}, fully-connected architectures \cite{grais2016combining},  U-Nets~\cite{jansson2017singing, doire2019interleaved}, GANs~\cite{stoller2018adversarial, choi2017singing}, as well as combinations of the above~\cite{uhlich2017improving, kavalerov2019universal}. Some research works suggested the estimation of each source separately with a dedicated network \cite{pritish2017deepconv, stoter19}, while other approaches  employed one-to-many encoder-decoder networks with a shared encoder and one decoder per source~\cite{doire2019interleaved}. Overall, the use of an individual network for each source seems to provide a better performance but it comes at the cost of increased training time.

There have been diverse proposals for loss functions, which include $L_2$-distance \cite{pritish2017deepconv, uhlich2017improving}, and $L_1$-distance \cite{jansson2017singing, doire2019interleaved} on estimated spectrograms, $L_2$-distance on ratio and binary masks \cite{grais2016combining}, $L_1$-distance on ratio masks \cite{gao2019co}, binary cross entropy on binary masks \cite{zhao2018soundofpixels, zhao2019soundofmotions}, as well as negative Scale-Invariant Signal to Distortion Ratio (SI-SDR) \cite{le2019sdr, luo2018tasnet} and Signal to Noise Ratio (SNR) \cite{kavalerov2019universal} as objective functions.


\subsection{Audio-visual approaches and source separation}

\subsubsection{Audio-visual model-based methods} 
Audio and visual information are related, and we can often see or imagine the source of origin for every particular sound. In the real world, they have a causal relation. In addition, the study of some misattribution effects (i.e. ventriloquism effect, see \cite{hershey2000audio} and references therein) has shown that people tend to relate audio and visual events if they happen simultaneously. Having this in mind, a correlation approach for source localisation was proposed as early as in 2000 \cite{hershey2000audio}. It consisted of calculating intensity changes in audio and video and computing correlations between audio and every pixel in a sequence of frames. The authors showed that the method can successfully identify the speaking person at every time frame in videos of two people speaking in turns.   


Thinking along the same line, Kidron~et~al. presented a method that detects pixels associated with a sound source while filtering out other dynamic pixels~\cite{kidron2005pixels}. The method used a refined version of canonical correlation analysis and, in contrast to previous studies, which mostly focused on speech applications, it could handle different types of sounding sources, not only people speaking but also musical instruments. The authors also discussed the \textit{chorus ambiguity} phenomenon when several people sing in synchrony, and in this particular case they accepted the detection of any of the faces as a successful result. The main concern raised by the authors is the extreme \textit{locality} of the pixel regions associated with an audio event which they overcome by introducing a sparsity constraint. That work was further extended in \cite{barzelay2007harmony}, incorporating temporal information for matching visual and audio onsets. 

A more recent work which focuses solely on chamber musical performances \cite{li2017see}, explored the association of musical scores with their spatio-temporal visual locations in video recordings. First, the authors performed audio-score alignment based on chroma features and Dynamic Time Warping, therefore automatically obtaining video-score alignment. Next, they used  optical flow to compute bow strokes motion velocities and correlated  them with audio onsets. The video analysis consisted of fitting a Gaussian Mixture Model for player detection and computing a histogram of motion magnitudes for fine-grained localisation of a high-motion region.


Parekh~et~al. \cite{parekh2018guiding} looked for sparse motion patterns which are similar to audio activation matrices obtained with Non-negative Matrix Factorization (NMF). In particular, from the visual modality, the authors computed frame-wise average magnitude velocities of clustered motion trajectories. Then, a linear transformation which transforms the motion velocity matrix into the spectral activation matrix was used to constrain the non-negative least square cost function together with a sparsity constraint. Both NMF and the audio-motion transformation were jointly optimised. The results showed a noticeable drop in signal-to-distortion ratio (SDR) when going from duos to quartets (from 7.14dB to 0.67dB for the best method while using soft masks for reconstruction). As in \cite{li2017see}, the performance of the method proposed in \cite{parekh2018guiding} decreases when separating sounds of the same instrument while addressing this problem for the first time. Interestingly, the authors only focused  on the motion component of videos ignoring other visual characteristics such as shape, color, and texture.

\subsubsection{Audio-visual deep learning methods} 
With the outbreak of deep learning techniques, the field of audio-visual learning has received a significant boost, especially the problems formulated in unsupervised and self-supervised manners. Along this line of research there have been some works focused on representation learning with further applications in audio classification, action recognition and source localisation \cite{aytar2016soundnet, arandjelovic2017look, arandjelovic2018objects, senocak2019learning, korbar2018cooperative, gao2019listen, parekh2019weakly, Liu2019Weakly}. Most of them combined features from two-stream networks (one sub-network for the audio and another one for the visual modality) either by concatenating them or by having an additional attention module. Some of them employed time synchrony for the samples of the same video \cite{owens2018audio, korbar2018cooperative}, while others learnt to extract features by identifying if the audio sample corresponded to a given visual data \cite{senocak2019learning, korbar2018cooperative, aytar2016soundnet}. More recent work also focused on the usage of audio for distilling redundant visual information to reduce computational costs~\cite{gao2019listen}.

Different objective functions such as cross-entropy \cite{arandjelovic2017look, arandjelovic2018objects}, Kullback–Leibler divergence \cite{aytar2016soundnet, gao2019listen}, contrastive \cite{korbar2018cooperative} or triplet \cite{senocak2019learning} losses were exploited in audio-visual deep learning. Distinctively, Korbar~et~al. \cite{korbar2018cooperative} used curriculum learning by first training the network with easy examples (correspondence was defined as being sampled from the same video) and then with hard/superhard examples (correspondence was defined as time-synchrony with/without time shift within the same video).

At the same time, the field of visually assisted source separation has emerged \cite{ephrat2018looking, owens2018audio, lu2019audio, parekh2019identify, xu2019recursive}, in particular, with explicit focus on musical data \cite{zhao2018soundofpixels, zhao2019soundofmotions, gao2018learning, gao2019co, xu2019recursive}.
Starting with capturing only visual appearance features  \cite{zhao2018soundofpixels, gao2018learning, gao2019co, xu2019recursive} there is a shift towards capturing and integrating motion data \cite{zhao2019soundofmotions}. 

To combine the data obtained from different modalities, commonly used approaches include late fusion \cite{parekh2019identify}, conditioning at the bottleneck via tile-and-multiply \cite{gao2019co}, concatenation \cite{ephrat2018looking}, attention mechanism \cite{zhao2018soundofpixels, zhao2019soundofmotions}, and Feature-wise Linear Modulation (FiLM) conditioning \cite{dumoulin2018featurewise, zhao2019soundofmotions} (more details in Section \ref{ss:conditioning}). In the present work, we also analyse different ways to combine audio and visual information and extend prior work for multiple and variable number of sources in the mixture.

It's also worth noting that two antecedent works in audio-visual source separation explored approaches that separate one source at a time, which can be applied for estimating multiple and variable number of sources \cite{gao2019co, xu2019recursive}. However, they have only been trained on \textit{artificial} mixtures of up to 4 sources and \textit{real} mixtures of 2 sources. The separation enhancement scheme proposed in \cite{xu2019recursive} consisted of extracting one source at a time from a residual audio mixture while considering maximum visual energy at every step, which follows the idea proposed in \cite{kavalerov2019universal}. Authors trained the network with mixtures of 2 and 3 instruments, and tested it on mixtures of up to 5 instruments.

Concurrently, the idea of co-separation has been proposed in \cite{gao2019co}. The method consisted of guiding source separation by integrating visual features of a detected musical instrument at the bottleneck of the primary U-Net, while the training was done using mix-and-separate approach with a combination of separation and consistency losses. The latter is defined as a cross-entropy loss between ground truth instrument labels and the predictions obtained with an additional classifier on the preliminary separated sources.

\subsection{Conditioned source separation}\label{ss:conditioning}

In the previous section we reviewed existing research in source separation combining information from visual and audio modalities. This approach can be reformulated as audio source separation \textit{conditioned} on visual information. We observe that, while there are several strategies of data fusion (i.e. concatenation or co-processing), another possibility is to modulate activations of a primary audio network by a context vector extracted from another modality. This technique is known as \textbf{F}eature-w\textbf{i}se \textbf{L}inear \textbf{M}odulation (\textbf{FiLM}) \cite{dumoulin2018featurewise}. The conceptual idea of FiLM conditioning is simple: it takes a set of learned features and scale and shift them accordingly to a context vector. Scaling and shifting parameters $ (\gamma, \beta) $ are learned based on an input context vector $\textbf{c}$ by an arbitrary function $f$ which is called FiLM-generator:

\begin{equation} 
(\gamma, \beta) = f(\textbf{c}). 
\end{equation}

The learned parameters modulate a neural network's activations $ F_i $  via a feature-wise affine transformation with $ i $ the feature map index:

\begin{equation}
FiLM(F_{i}|\gamma_{i}, \beta_{i}) = \gamma_{i}F_{i} + \beta_{i}.
\end{equation}

Other studies considered \textit{weak conditioning} in source separation using only labels of target sources \cite{meseguer2019conditioned, slizovskaia2019end} in contrast to \textit{strong conditioning} where the context vector could be available frame-wise \cite{tzinis2019improving, schulze2019weakly}. The employed weak label conditioning techniques include FiLM \cite{meseguer2019conditioned} and tile-and-multiply \cite{slizovskaia2019end}. For strong conditioning, a binary vocal activity vector and vocals magnitude vector have been used for singing voice separation with attention mechanism \cite{schulze2019weakly}.  

Later, the idea has been explored in the context of universal source separation with conditioning on classification embeddings \cite{tzinis2019improving}. First, the method extracts the context embeddings with the classification network, then upsample and normalise them, which is followed by conditioning of the primary source separation network either by concatenation with the network's activations or gating the activations by the embeddings. Another work goes along this line and trained a source separation model based solely on weak labels \cite{pishdadian2019finding}. The method consists of training a classifier network and using the classification loss (with an additional constraint for the estimated sources to sum to the mixture) as the objective function for separation. 

We find various strategies to integrate side information, and different modules of the network being conditioned. However, most of the studies injected the context vector at the bottleneck of encoder-decoder architecture with a rare exception of early fusion in \cite{tzinis2019improving}. The same authors \cite{tzinis2019improving} reported that integration of the context vector at every layer of the primary network leads to overfitting.



\section{Approach}\label{ss:approach}

\begin{figure*}[ht]
    \centering
    \includegraphics[width=0.8\linewidth]{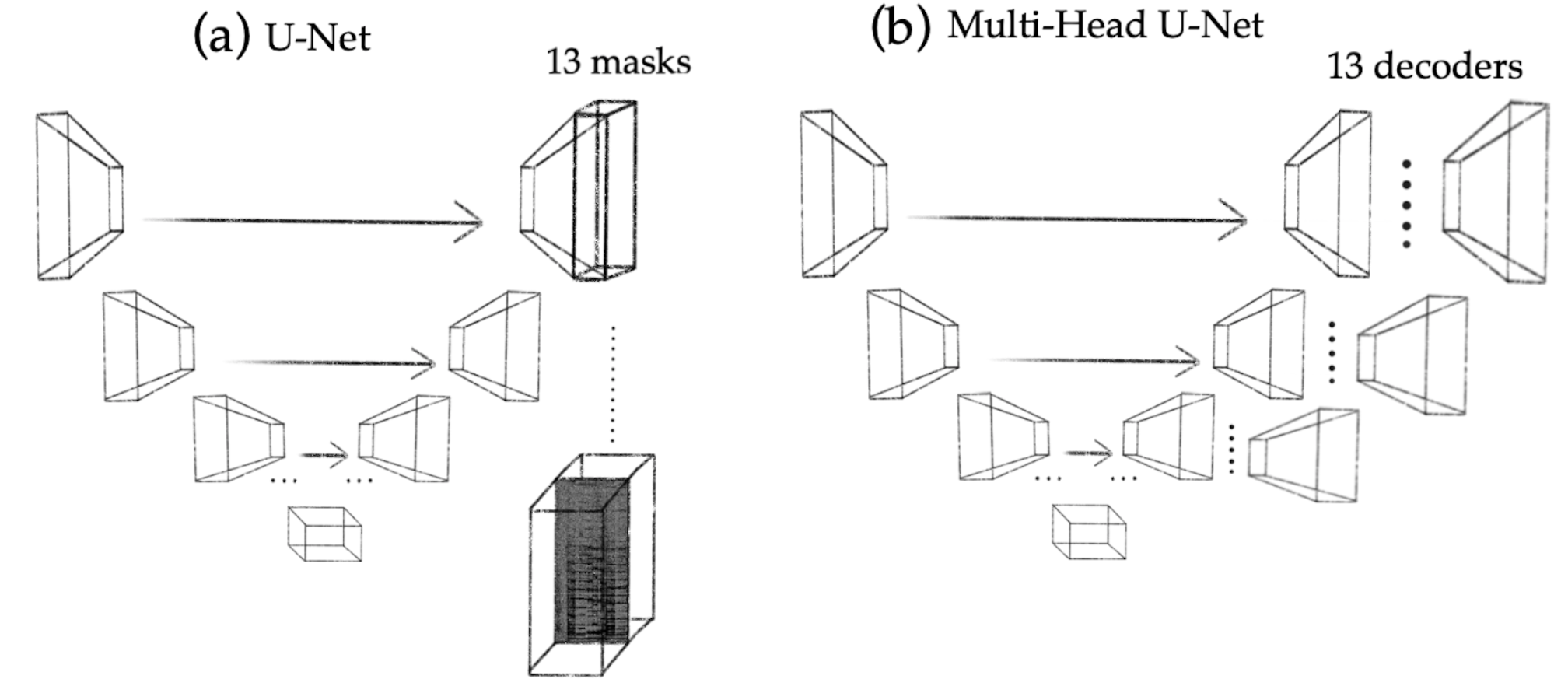}\\
    \includegraphics[width=0.9\linewidth]{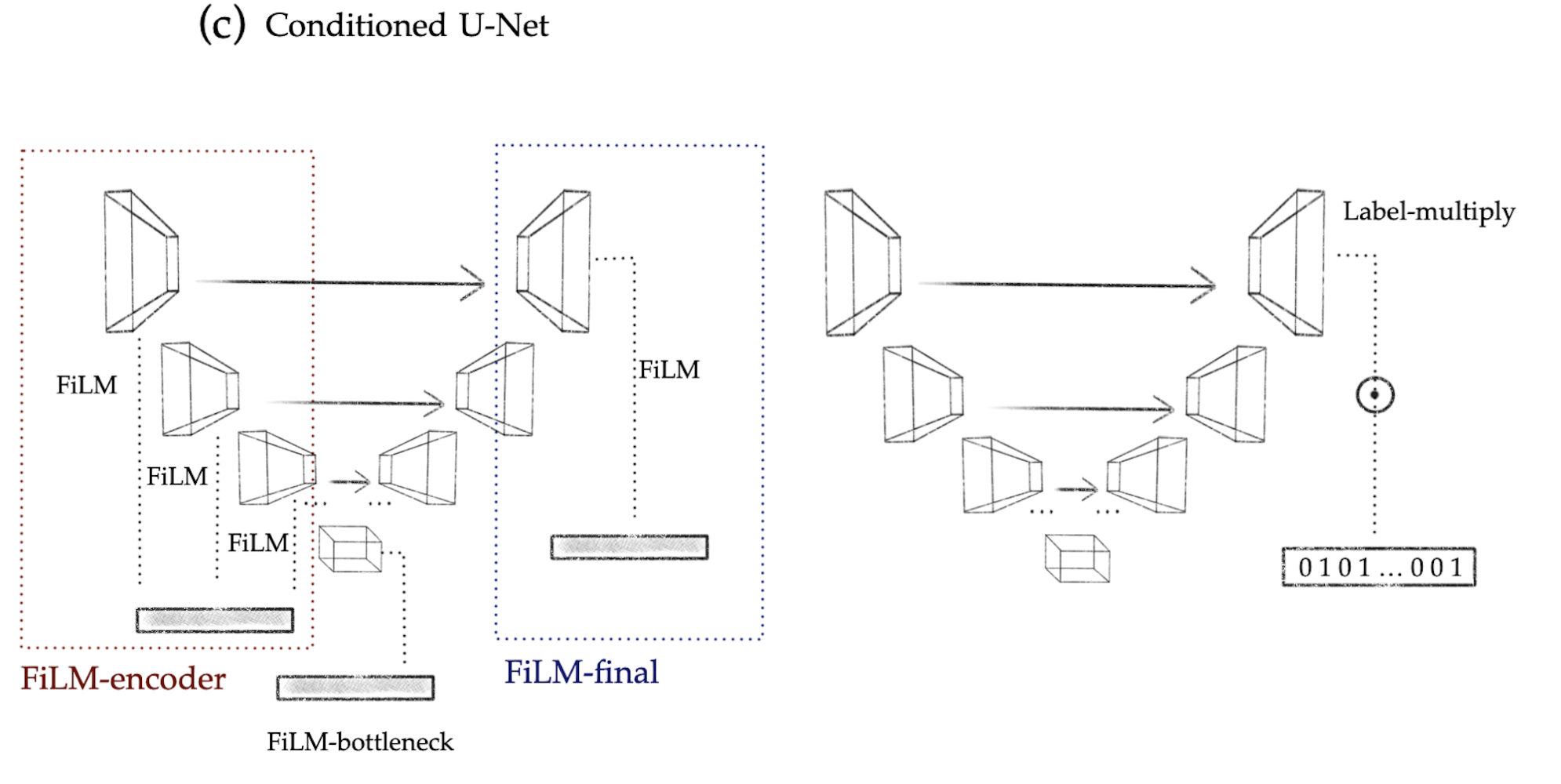}
    \caption{Summary of architectures and conditioning types. There are two variants for the source separation architecture: (a) U-Net which outputs 13 masks at the last upconvolutional layer, (b) Multi-Head U-Net with one shared encoder and 13 specialized decoders which output one mask each. (c) There are several choices for U-Net conditioning: three types of FiLM conditioning and multiplicative conditioning of the output masks.}
    \label{fig:unet-all-2}
\end{figure*}

In this work, we study the effect of integrating two types of context information at different locations of the network, while keeping the architecture fixed and simple.
As context information we explore: i) labels, that indicate the presence or absence of each type of instrument in the mixture, and ii) visual features.

To the best of our knowledge, there is a lack of public datasets in the format of audio-visual music recordings with ground-truth individual sources. Thus, we use a mix-and-separate approach for training, such that every mixture is generated on the fly and, therefore, unique. To create a mixture, we take the following steps: (1) we sample an arbitrary subset of instruments; (2) we subsequently pick a random segment from one of the audios of each instrument category; and (3) we scale values in each segment to the $[-1, 1]$ range in order to adjust loudness, then sum the segments and normalize the sum by the number of instruments to avoid clipping.
Given a magnitude spectrogram of the mixture, our network learns to predict $K$ real-valued masks $\hat{M}_i$, one mask per potential instrument present in the mixture (we use $K=13$ different instruments in our experiments, see Figure \ref{fig:unet-all-2}a U-Net and Figure \ref{fig:diagram}). Each output mask is associated to a certain type of instrument, and their order is fixed. 

\subsection{U-Net and Multi-Head U-Net baselines}\label{ss:approach:baseline}

We adopt two simple U-Net models as the baseline architectures, given that U-Net has been extensively used and demonstrated good performance \cite{jansson2017singing, doire2019interleaved, zhao2018soundofpixels, zhao2019soundofmotions, gao2019co}. As the focus of this work is on studying the effect of different types of conditioning, we leave for future research the analysis of different source separation networks. 

U-Net\cite{unet} is an encoder-decoder architecture with \textit{skip connections} such that activations of every $i$-{th} layer of the encoder are concatenated with activations of $(N-i)$-{th} layer of the decoder, which can be considered as a light form of conditioning by itself. Following \cite{zhao2018soundofpixels, zhao2019soundofmotions}, we have chosen one of the architectures they propose and set the number of layers to $N=6$. We employ two variants of the architecture, namely: (a) a baseline U-Net architecture as pictured in Figure \ref{fig:unet-all-2}a which outputs 13 masks after the last upconvolutional layer, and (b) Multi-Head U-Net (MHU-Net) \cite{doire2019interleaved} as pictured in Figure \ref{fig:unet-all-2}b which has a single shared encoder and 13 decoders, where each dedicated decoder yields a mask for its corresponding instrument.

Audio is resampled at 11025 Hz before preprocessing in order to speed up computation, based on \cite{zhao2018soundofpixels}, where the authors argue that most perceptually important frequencies of instruments are preserved and audio quality is only slightly reduced. We use Hann window, and the STFT is computed for every segment of approx. 6 seconds  (65535 audio samples) with window size of 1022 (about 93ms) and hop size of 256, which results in a matrix of $ 512 \times 256 $ STFT bins. All those parameters are taken from \cite{zhao2018soundofpixels} and \cite{gao2019co} (the same hop length and approx. the same window length in time were also used in \cite{xu2019recursive}). Next, we study a few preprocessing strategies over the STFT representation, including linear and log-sampled frequency scale for the STFT, as well as log-scale and dB-scale with normalisation for the STFT magnitude values as discussed in Section \ref{ablation}. 

The choice of loss functions is dependent on the type of mask. 
For binary masks, we compute binary cross entropy (BCE) loss:
\begin{equation}\label{eq:BCE}
\begin{split}
 \mathcal{L}^{b} = &- \sum_{i=1}^{K} \sum_{(\tau,\omega)} (\lambda \, M_{i}^{ib}(\tau, \omega) \log(\hat{M}_{i}(\tau,\omega)) \\ &+ (1 - M_{i}^{ib}(\tau,\omega)) \log ( 1 - \hat{M}_{i}(\tau,\omega))),
 \end{split}
\end{equation}
where $M_i^{ib}(\tau, \omega)$ and $\hat{M}_i(\tau, \omega)$ represent, respectively, ground truth and predicted binary mask values at frequency bin $\omega$ and time frame $\tau$, $K$ is the maximum number of sources, 
and $\lambda$ is a positive (fixed) weight which is used to compensate for the class imbalance in the mask values.


For ratio masks we employ $L_2$ loss:
\begin{equation}\label{eq:L2}
    \mathcal{L}^{r} = \sum_{i=1}^{K} 
    \|M_i^{ir} - \hat{M}_i\|_2^2,
\end{equation}
where $M_i^{ir}$ and $\hat{M}_i$ denote, respectively, ground truth and predicted ratio masks.

\subsection{Conditioned U-Net} \label{ss:approach:cunet}

\begin{figure*}[ht]
    \centering
  \subfloat[The main separation network receives as input the spectogram of the mixture and a context vector. It outputs $K$ masks. \label{1a}]{%
       \includegraphics[width=0.45\linewidth]{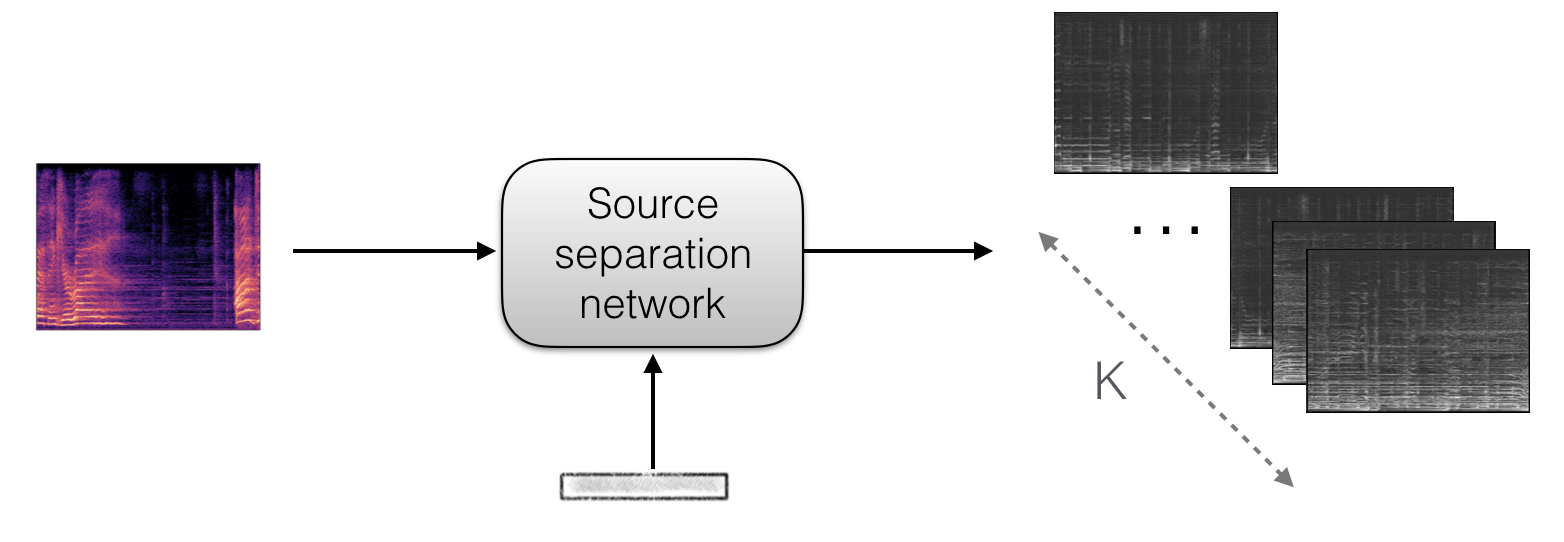}}
    \hfill
  \subfloat[Types of context vectors: binary (left), visual (center), and visual-motion (right). The inputs in each case are: binary vector (left), one frame per instrument (center), a set of frames per instrument (right).\label{1b}]{%
        \includegraphics[width=0.45\linewidth]{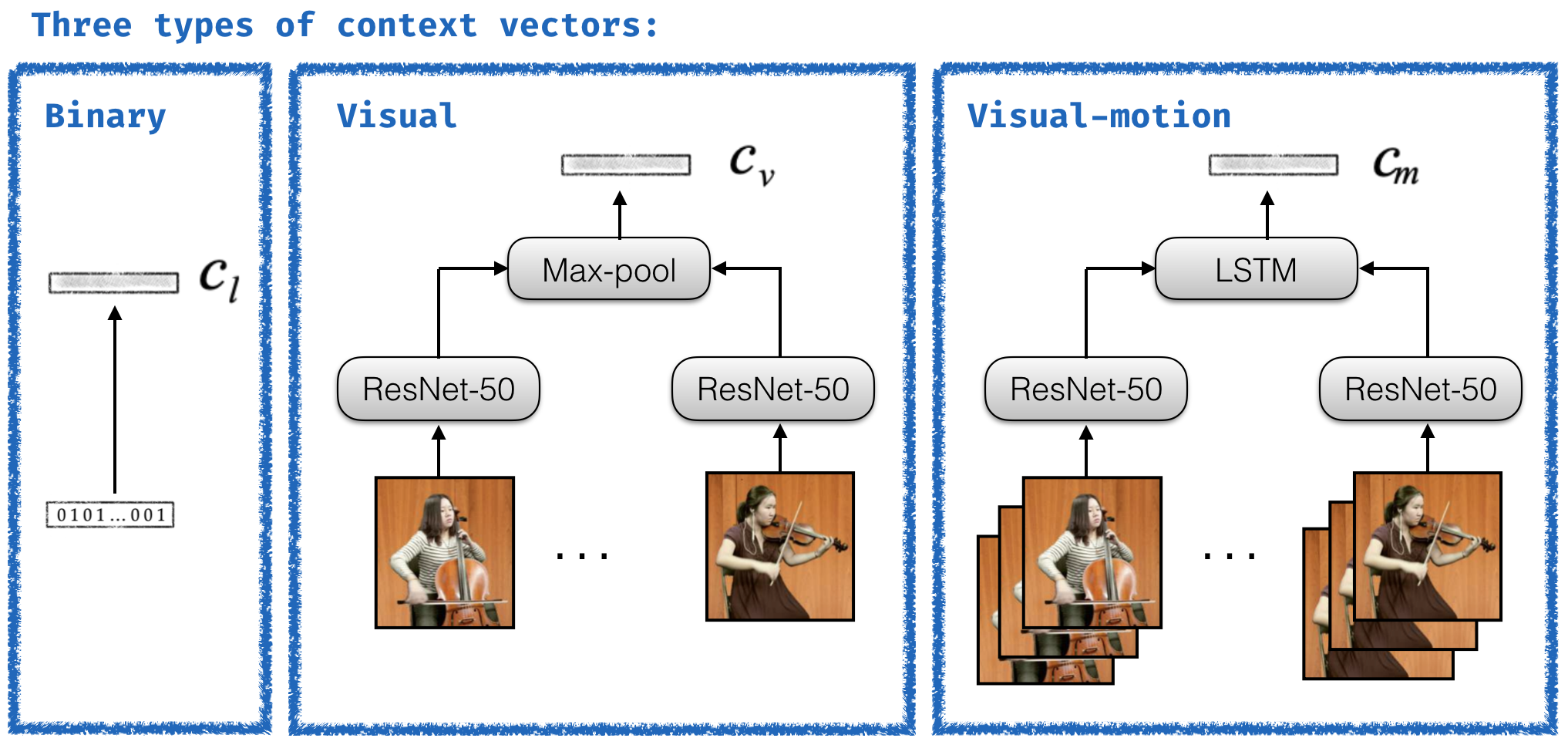}}
    \caption{Diagrams showing the inputs, outputs and types of conditioning vectors.}
    \label{fig:diagram}
\end{figure*}

In this section we describe the conditioning strategies and the types of context data (Figure \ref{1b}) which we use in our Conditioned U-Net architecture (Figure \ref{fig:unet-all-2}c).

\subsubsection{Weak label conditioning}\label{ss:approach:cunet:labels}

We study \textit{weak conditioning} for source separation which means that instrument labels are available at the level of individual recordings. They indicate the presence or absence of each instrument in the mix, which is encoded in a binary indicator vector $ \textbf{c}_l \in \{0,1\}^{K} $  where $K$ is the total number of instrument classes considered. In all network variants, the output is a set of $K$ masks (Figure \ref{1a}), and we denote by $\hat{M}_i$  the mask associated to the $i$-{th} instrument type.

We use $ \textbf{c}_l $ as a conditioning context vector and compare three types of FiLM conditioning: introduced (1) at the bottleneck, (2) at all encoder layers, and (3) at the final decoder layer (see Figure \ref{fig:unet-all-2}c). More formally, for each layer $j$ (no matter at which location the FiLM is used) we have activations or embeddings: $ \boldsymbol{a}^{(j)} $, and the conditioning is as follows:
\begin{equation}
  (\gamma_j, \beta_j) = f_j(\boldsymbol{c}_l)
\end{equation}
\begin{equation}
  \boldsymbol{\hat{a}}^{(j)} = \gamma_j \boldsymbol{a}^{(j)} + \beta_j
\end{equation}
where $f_j$ is a single fully connected layer.

Furthermore, we explore simple multiplicative conditioning with the binary indicator vector: 
\begin{equation}
  \hat{M}_i = \boldsymbol{c}_{l}[i] \Tilde{M}_i, 
\end{equation}
where $ \textbf{c}_l[k]$ is the $i$-th component of the context vector and $ \Tilde{M}_i$ is the $i$-th preliminary mask as predicted by (MH)U-Net.


\subsubsection{Visual conditioning} \label{ss:approach:cunet:visuals}

In the case of visually-informed source separation, we consider both static characteristics and motion-aware conditioning. Nonetheless, we would like to note that learning temporal information from videos is a challenging task which is still under research. Therefore, visually-informed methods mostly use a single frame for conditioning \cite{zhao2018soundofpixels, gao2019co, gao2018learning}, with some exception of dense trajectories \cite{li2017see}, and deep-learned dense trajectories \cite{zhao2019soundofmotions}.

Alike \cite{parekh2018guiding, gao2019co} we assume that rough spatial location of each source is given (e.g. it can be obtained by a segmentation or human detection algorithm). Keeping this assumption in mind, we use uncropped frames from individual videos for training and evaluation (since videos in the Solos dataset contain a single instrument). In a real life scenario (e.g. for testing) we use a bounding box around every player. In the experiments with the URMP dataset we manually annotated the bounding boxes.

For visual context conditioning, we take a single video frame corresponding to the beginning of the audio source sample. We use a pretrained ResNet-50 \cite{he2016resnet} to extract a visual feature vector of size 2048 for every present source, and then concatenate them, obtaining a visual context vector $ \textbf{c}_v $ of size $ K^{\prime} \times 2048 $ where $ K^{\prime} $ is the maximum number of sources in the mixture ($K^{\prime} \leq K$). The context vector for the unavailable sources is set to all zeros. To obtain a final conditioning vector, we first pass the visual feature vectors through an adaptive max pooling and obtain a conditioning of size $ K^{\prime} \times 1024$. Then, we take a per-channel maximum over all $ K^{\prime}$ frames and obtain a single 1024-D conditioning vector. As for the case of weak label conditioning, we compare three alternatives for the FiLM conditioning (see Figure \ref{fig:unet-all-2}c).
Additionally, we conduct an experiment which is similar to label multiplicative conditioning, but instead of using labels we use a vector of instrument probabilities. Class probabilities are obtained by adding an additional fully-connected layer with a softmax activation after the context vectors. This experiment is denoted later in Section \ref{ss:results:visual} as Final-multiply. 

For visual-motion conditioning, we first extract visual feature vectors with the pretrained ResNet-50 at a fixed frame-rate within the video segment corresponding to the selected 6s segment of audio. We then pass the obtained sequence of vectors through a small uni-directional LSTM network with 1024 hidden units, as in \cite{gao2019listen}, with the aim of capturing motion characteristics while keeping visual information. We take the last LSTM hidden state resulting in a motion context vector $ \textbf{c}_m $ of size $1024 $.  Due to the large computational cost, and based on the results of the ablation study (Section \ref{ss:results:visual}), we only report this approach with FiLM conditioning at the bottleneck and the final layer of audio \mbox{U-Net}. Alternatively, we analyse another strategy which consists of directly max-pooling visual features extracted from all frames into a single conditioning vector.

\section{Experiments}\label{ss:experiments}
In what follows, we thoroughly evaluate the proposed method on various setups. In particular, we compare the  different conditioned networks with respect to several performance metrics.

\subsection{Dataset}\label{ss:datasets}

In our experiments we use two multimodal datasets of musical performances: the recently introduced Solos dataset \cite{solos} for training and evaluation, and the URMP dataset \cite{urmpdataset} for testing (Table \ref{tab:datasets} provides a summary of both datasets). 

\textbf{The URMP dataset}\footnote{\url{http://www2.ece.rochester.edu/projects/air/projects/URMP.html}} consists of 44 arrangements (of which 11 are duets, 12 are trios, 14 are quartets, and 7 are quintets) with a total duration of 1h 14min. Each source track was recorded separately with an external coordination, and the final mixes were assembled afterwards.   
The instrumentation is a typical one for chamber and orchestral music, and includes families of instruments such as strings (violin, viola, cello and double bass), woodwinds (flute, oboe, clarinet, bassoon, saxophone), and brass (trumpet, horn, trombone, tuba). The dataset is constructed to reflect the complexity of the musical world where the same instrument within a section can appear more than once.  
For each
piece, the dataset provides the musical score in MIDI format,
the high-quality individual instrument audio recordings and
the videos of the assembled pieces. For that, all individual visual contents were composited in a
single video with a common background where all players
were arranged at the same level from left to right.

As in this work we only tackle the problem of separating sources of different instruments, we mix source tracks of the same instrument within the same piece and consider the resulting mix as a single source. For example, for a string quartet (which consists of 2 violins, viola and cello), we mix the two source tracks of the violins which results in a corresponding ``trio" where two violins are considered as a single source. Also, we remove four pieces ({02\_Sonata\_vn\_vn}, {04\_Allegro\_fl\_fl}, {05\_Entertainer\_tpt\_tpt}, {06\_Entertainer\_sax\_sax}) from the dataset as they are duets of the same instrument. After this preprocessing, we are left with 12 duets, 20 trios and 8 quartets in the final set.  

\textbf{The Solos dataset}\footnote{\url{https://www.juanmontesinos.com/Solos/}} consists of 755 YouTube videos of solo musical performances of the same 13 instruments categories as of the URMP dataset. 
It has a total duration of about 66 hours. A major part of the dataset are audition performances which ensures, together with manual and semi-automatic checking, a good quality of audio and video. We use this data with a mix-and-separate strategy and complement it with an evaluation of the trained models in the URMP dataset, which allows proper evaluation on real-wold mixtures.
Although most of the previous works on audio-visual learning-based source separation have been trained on the MUSIC dataset introduced in \cite{zhao2018soundofpixels}, it lacks 6 out of the 13 instrument classes present in the URMP dataset (namely, viola, double bass, oboe, bassoon, horn, trombone). That is why we choose to train on Solos.

\begin{table*}[]
\centering
\begin{tabular}{l|cccc|cc}
  & \multicolumn{4}{c}{URMP} & \multicolumn{2}{c}{Solos}\\
Category  & \# Recordings  & \# Pieces  & (\# duets, \# trios, \# quartets) & Mean duration & \# Recordings & Mean duration \\ \hline
Violin                  & 11 & 20 & (5, 12, 3) & 1:51  & 66            & 6:16          \\
Viola                   & 7 & 12 & (1, 8, 3) & 1:52   & 55            & 5:31          \\
Cello                   & 10 & 11 & (3, 7, 1) & 1:52  & 134           & 7:21          \\
DoubleBass              & 3 & 4 & (0, 1, 2) & 2:28    & 58            & 8:53          \\ \hdashline

Flute                   & 8 & 11 & (3, 4, 4) & 1:50   & 48            & 4:00           \\
Oboe                    & 3 & 6 & (1, 2, 3) & 1:58   & 53            & 5:45           \\
Clarinet                & 8 & 10 & (2, 5, 3) & 1:44   & 49            & 3:23          \\
Bassoon                 & 2 & 3 & (0, 0, 3) & 1:25    & 56            & 5:08          \\
Saxophone               & 8 & 9 & (2, 5, 2) & 1:31    & 45            & 2:42          \\ \hdashline

Trumpet                 & 11 & 12 & (5, 5, 2) & 2:07   & 50            & 1:14          \\
Horn                    & 5 & 5 & (0, 3, 2) & 2:02    & 50            & 5:11          \\
Trombone                & 8 & 9 & (2, 5, 2) & 2:01    & 50            & 5:03          \\ 
Tuba                    & 3 & 5 & (0, 3, 2) & 1:47    & 41            & 2:49          \\ \hdashline

{\bf TOTAL}             & {\bf 87} & {\bf 40} & {\bf (12, 20, 8)} & {\bf 1:53}  & {\bf 755}     & {\bf 5:16}    
\end{tabular}
\caption{Statistics of the datasets used: URMP and Solos. Notice that, as commented in Section \ref{ss:datasets}, we consider different instances of the same instrument as the same source, that is why the resulting number of pieces, duos, trios, and quartets varies from the original dataset (e.g. there are no quintets left since in all of them at least one of the instruments is repeated).}
\label{tab:datasets}
\end{table*}

\subsection{Metrics} 

Several studies indicate that widely-adopted source separation metrics such as signal to distortion ratio (SDR), signal to inference ratio (SIR), and signal to artifacts ratio (SAR) \cite{vincent2006sdr} do not always agree with human perception \cite{le2019sdr, kilgour2019frechet, zhao2018soundofpixels, gao2019co}. Moreover, as brought out in \cite{le2019sdr}, an increment of noise or interferences in the separated source produces an increment of the SAR value. Recently, \textit{scale-invariant} and \textit{scale-dependent} SDR (SI-SDR, SD-SDR)  have been proposed \cite{le2019sdr} as better alternative metrics. For completeness, we report all these measures.

Unlike previous works \cite{gao2019co, xu2019recursive, zhao2018soundofpixels, zhao2019soundofmotions}, our method produces \textit{sparse} outputs since many predicted sources are expected to be silent. However, all the above metrics are ill-defined for silent sources and targets. To address this issue, we also compute cumulative \textit{predicted energy at silence} (PES) as proposed in \cite{stoller2018jointly} and later refined in \cite{schulze2019weakly}. For SI-SDR and SD-SDR larger values indicate better performance, for PES smaller values indicate better performance. For numerical stability of log function, in our implementation we add a small constant $\epsilon = 10^{-9}$ which results in a lower boundary of the metrics to be $-80$ dB.

\subsection{Training and implementation details}\label{ss:tr_details}

Our U-Net is composed of six blocks in the encoder and six blocks in the decoder. Each encoder block consists of a convolutional layer followed by batch normalisation with an optional conditioning layer (for FiLM-encoder conditioning), and LeakyReLU nonlinearity. \textcolor{black}{We use convolutions of $4 \times 4$ with stride two and padding one. The number of filters in the initial layer is set to 16, and then doubled for each consecutive layer until it reaches 512 in the 6th block.} A decoder block consists of a bilinear upsampling layer \textcolor{black}{with a scale factor of two}, a convolutional layer, batch normalisation, ReLU nonlinearity, and a dropout layer \textcolor{black}{with $p = 0.2$. The convolutional filters in the decoder blocks are of size $3 \times 3$ with stride one and padding one.}

The network is trained for 1000 epochs with a batch size of 32, Adam optimizer, an initial learning rate of $10^{-5}$ which is halved after 25k iterations with no improvement on the validation set.

We have also analysed the use of a curriculum learning strategy. It consists of starting the training with only mixtures of 2 sources, and gradually increasing the maximum number of sources up to 7. The increment is carried out if the validation loss does not decrease for 10k iterations. Some examples where curriculum learning has proven to be beneficial for training audio-visual networks are \cite{zhao2019soundofmotions} for source separation, and \cite{korbar2018cooperative} for audio-visual temporal synchronisation.

For training and evaluation we utilize the mix-and-separate procedure by creating artificial mixtures from individual videos of Solos. Every training sample has an arbitrary number of sources \textcolor{black}{in the mixture, with an upper bound of the maximum number of sources to appear simultaneously is set to 7}. For testing, we use real mixtures from the URMP dataset which have at most 4 simultaneously sounding sources.

\textcolor{black}{The implementation takes advantage of various Python libraries and uses PyTorch as a deep learning framework. The code is made publicly available\footnote{\url{https://github.com/Veleslavia/conditioned-u-net/tree/master/code}}.}

\subsection{Baseline ablation study}\label{ablation}


\begin{table*}[ht]
    \centering
    \begin{tabular}{c||c|c|c|c|c|c|c|c}
        Method & Exp. ID & W & SI-SDR $\uparrow$ & SD-SDR $\uparrow$ & PES $\downarrow$ & SDR $\uparrow$ & SIR $\uparrow$ & SAR $\uparrow$ \\ \hline 
        IRM & U & 0 & $ 13.1\pm5.4 $ & $ 12.7\pm6.4 $  & n/a & $ 11.8\pm4.3 $ & $ 19.9\pm5.6 $ & $ 13\pm4.3 $ \\ 
        input mix & L & 0 & $-3.7\pm5.7$ & $-3.7\pm5.7$ & $18.2\pm4.2$ & $ -3.5\pm4.8 $ & $ -3.2\pm4.9 $ & $ 18.1\pm11.2 $ \\ \hline
        No filtering & 1 & 0 & $  -2.1\pm 6.2 $ & $ -15.2\pm 10.0 $ & $ -8.1\pm 5.9 $ & $ -0.8\pm5.8 $ & $ 0.6\pm6.8 $ & $ 10.9\pm3.2 $ \\
        Wiener (1 it.) & 1 & 1 & $ -3.9\pm 7.5 $ & $ -16.0\pm 16.1 $ & $ -15.1\pm 10.1 $ & $ -1\pm6.2 $ & $ 1.5\pm7.7 $ & $ 7.7\pm3.8 $ \\
        Wiener (2 it.) & 1 & 2 & $ -7.6\pm 10.7 $ & $ -21.4\pm 24.1 $ & $ -25.4\pm 16.8 $ & $ -2.6\pm7.1 $ & $ \mathbf{1.7\pm8.5} $ & $ 5\pm5.6 $ \\ \hline 
        Exp. 1 $+_{CL}$ & 2 & 2 & $ -8.2\pm 10.9 $ &  $ -24.1\pm 25.1  $ &  $ -26.2\pm 17.8  $ &  $ -3.1\pm7.1  $ & \dotuline{$ 1.4\pm8.5 $} & \dotuline{$ 4.6\pm5.7 $}  \\
        Exp. 1 $+_{BM}$ & 3 & 2 & \dotuline{$ -9.7\pm 15.2  $} & \dotuline{$ -21.6\pm 27.4 $} &  $ \mathbf{-36.2\pm 18.4}  $ &  $ -3.8\pm8  $ &  $ 0.1\pm7.1  $ &  $ 4.9\pm7.5  $  \\
        Exp. 1 $+_{noise}$ & 4 & 2 & \dotuline{$ -7.6\pm 9.2  $} &  $-21.7\pm 20.4  $ &  $ -21.7\pm 15.6 $ &  $ -3.1\pm6.9  $ &  $ 0.2\pm8.3  $ &  $ \mathbf{6.0\pm5.3} $ \\
        Exp. 1 $+_{log-normalise}$ & 5 & 2 & $ -10.1\pm 11.6 $ &  $ -25.3\pm 24.9  $ &  $ -27.0\pm 17.0  $ &  $ -3.9\pm7.6  $ &  $ 0.7\pm9.5  $ &  $ 4.6\pm6.1  $ \\ 
        Exp. 1 $+_{linear~STFT}$ & 6 & 2 &  \dotuline{$\mathbf{-7.4\pm 9.9} $} &  $ \mathbf{-17.5\pm 19.8} $ &  $ -24.4\pm 15.8  $ & \dotuline{$\mathbf{-2.3\pm6.9}$} & \dotuline{$ 1.5\pm8.4 $} & $ 5.7\pm5.2 $ \\
        Exp. 1 $+_{MHU-Net}$  & 7 & 2 & $ -8.1\pm9.1 $ & $ -23.7\pm20.4 $ & $ -20.9\pm15.8 $ & $ -3.3\pm6.4 $ &  $ 0.1\pm8.2 $ &  $ 5.8\pm4.5 $ \\
\end{tabular}
    \caption{Ablation studies results for the URMP dataset with no conditioning. The first two rows indicate two references: ideal ratio masks (IRM, U states for the upper bound), and the usage of the input mixture as a predicted source (input mix, L states for the lower bound). Experiment IDs are explained in Table \ref{tab:cunet_ablation_parameters}. The column W indicates the number of iterations of Wiener filtering used in post-processing. The best results among all variants with 2 iterations of Wiener filtering are highlighted in {\bf bold}. Results that are {\bf not statistically significant} ($ p < 0.05 $) w.r.t. Exp.~1 with 2 iterations of Wiener filtering are \dotuline{dotted}. CL denotes the usage of curriculum learning and BM states for binary masks.}
    \label{tab:baseline_urmp}
\end{table*}

In preparation for conditioned source separation analysis and to define the optimal hyperparameters of the baseline U-Net architecture as described in Section \ref{ss:approach:baseline}, we conduct a series of ablation experiments. We examine the following set of hyperparameters: (1) linear vs. log frequency scale for the STFT representation, (2) binary vs. ratio masks estimation, (3) data augmentation with normally-distributed noise, (4) log vs. dB-normalised scale for the STFT values, (5) the use of curriculum learning, and (6) the effectiveness of Multi-Head U-Net vs. vanilla U-Net.

\section{Results and Discussion}\label{ss:results}

\subsection{Ablation studies}\label{ss:exp_ablation}

We report the metrics obtained by unconditioned U-Net models in the ablation study in Table \ref{tab:baseline_urmp}, and the full list of hyperparameters is given in Appendix \ref{app:hyperparams}. The experiments can be matched by the experiment ID. We also provide two reference metrics, the upper bound separation quality (U) with ideal ratio masks (IRM), and the mixture metrics (L) which reproduce the input mixture at every possible output source.  
Results that are {\bf not} statistically significant w.r.t. Exp.~1 ($ p < 0.05 $) are dotted.

\begin{figure}
    \centering
    \includegraphics[width=\linewidth]{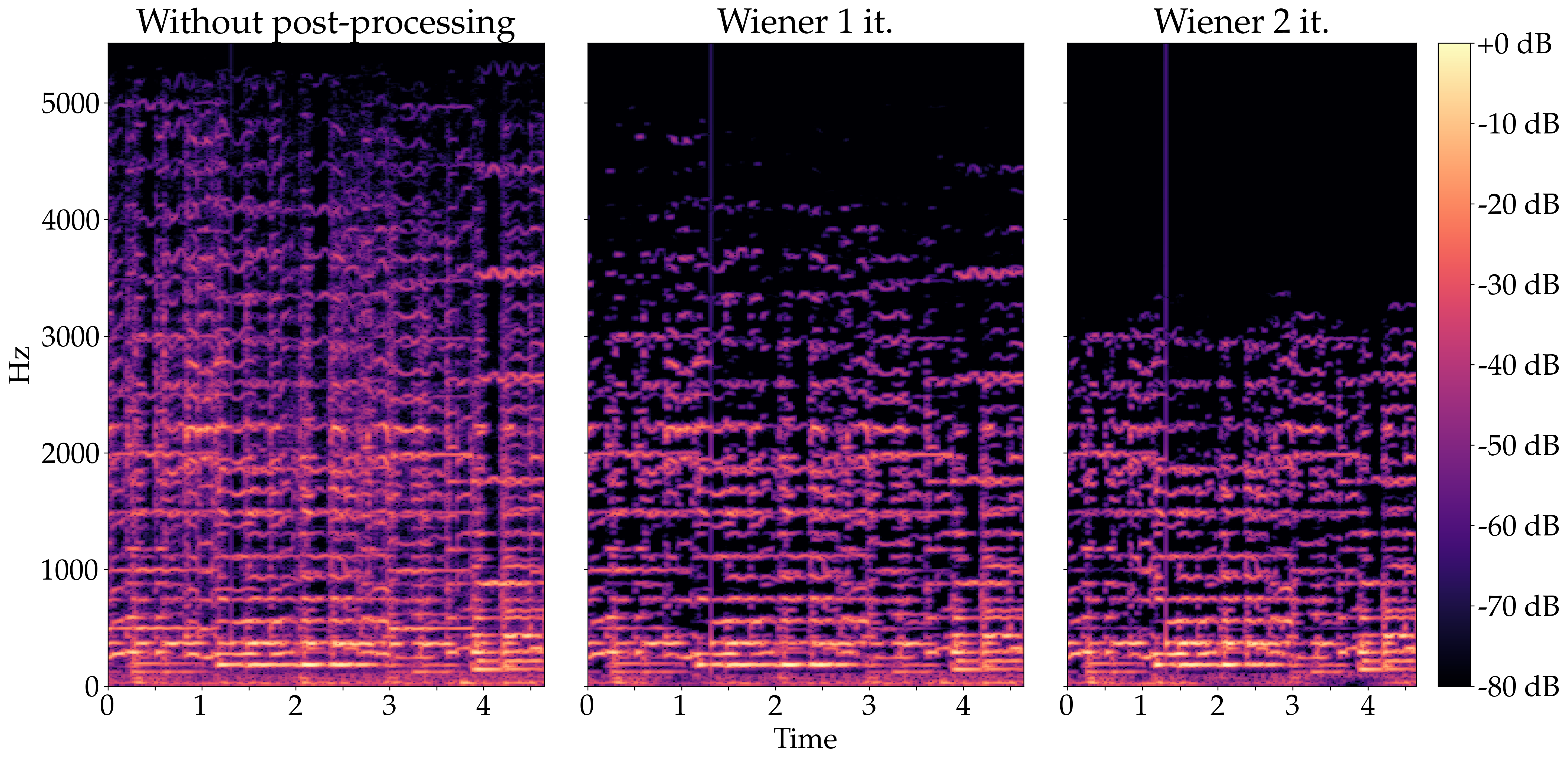}
    \caption{Effect of post-processing with 1 and 2 iterations of Wiener filtering for a test sample from Exp. 11 results. This is an example of separated cello with interferences from violin and viola voices. We can observe that, while audio quality is degrading and the separated part becomes more filtered, the interferences are also greatly reduced. We use linear STFT scale and visualise frequencies up to 5.5~kHz.}
    \label{fig:wiener_filtering}
\end{figure}

In practice, we noticed that, while training with ratio masks, the to-be-silent output sources eventually happen to be an original mixture with a lowered volume. Therefore, as it is common practice in audio source separation, we have applied a post-processing based on some iterations of the Wiener filter \cite{sigsep_norbert}. We indicate the number of iterations of Wiener filtering applied in the third column (W) of Table \ref{tab:baseline_urmp}.
The effect of the post-processing iterations can be observed: although  some of the metrics are degraded, the improvement in \mbox{PES} and \mbox{SIR} is considerable. 
The metric degradation can be seen mainly in Table \ref{tab:baseline_urmp}, in the second horizontal block. Figure \ref{fig:wiener_filtering} allows deeper analysis of quality changes, which is generic for post-processing in all experiments. In spite of the fact that the post-processed results sound more filtered and have artefacts, we can perceive a notable separation improvement when it comes to interferences. We believe that for multi-source separation with many silent source, the practical trade-off between quality degradation and separation improvement offered by post-processing is worth considering. In a real case scenario, the final choice would probably depend on the application. For example, for remixing it would be beneficial to preserve the original quality, but fewer interferences could be more desirable if the source separation is used as a preliminary step for automatic transcription. In this study we try to find a balance between different performance metrics and different perceptual aspects of separated audio. Therefore, in all following experiments we apply two iterations of Wiener post-processing.

As mentioned in Section \ref{ss:tr_details}, we have analysed the use of a curriculum learning (CL) strategy (Exp. 2) but as it can be observed it is only slightly beneficial in terms of \mbox{PES} while comparing with Exp. 1. We have also used binary masks instead of ratio masks (Exp. 3), the results are worse in general but the silent sources are better recovered which is resulted in lower \mbox{PES} (-10.8dB comparing to Exp. 1). 
Another aspect we have observed is that augmenting the input data with normally-distributed noise (Exp. 4) does not improve separation performance (except for SAR). However, we are of the view that the problem can benefit from other augmentation techniques, such as the ones described in \cite{uhlich2017improving, miron2017generating}. This is left for future research.

Results from Exp. 5 show that using normalised STFT is beneficial and Exp. 6 reveals that working with a linear-scale STFT gives overall better results. All measures in Exp. 6 have improved with respect to a log-scale STFT (Exp. 1) with the exception of PES and SIR which are barely degraded. In particular, the improvements in SD-SDR and SAR are statistically significant. 
Finally, the use of a multi-decoder architecture (Exp. 7), which  requires higher computational cost, does not improve the metrics (except for SAR).

In summary, the optimal parameters we found are dB-normalised and linear-scale STFT as input, a single decoder that predicts ratio masks, and Wiener post-processing. We have opted out of augmenting the input with normally-distributed noise and using curriculum learning. This configuration will be used in the following sections as baseline for the conditioning experiments.

\subsection{Conditioning on labels}\label{ss:results:weak}

We further study weak label conditioning of the single-decoder U-Net model. We provide results for 
four conditioning schemes as described in Section \ref{ss:approach:cunet:labels}. The summary of weak label conditioned source separation is shown in Table \ref{tab:weakly_conditioned_urmp}. 

We observe that the best performance in terms of \mbox{SI-SDR}, \mbox{SDR}, and \mbox{SAR} is obtained with FiLM conditioning in the encoder (Exp. 9, Table \ref{tab:weakly_conditioned_urmp}), but it also leads to high PES, even worse than in the case of no post-processing in the ablation study (Exp. 1, W $= 0$, Table \ref{tab:baseline_urmp}). On the other hand, the best results in terms of \mbox{SD-SDR} and \mbox{SIR} are achieved with the multiplicative conditioning on the output masks (Exp. 11, Table \ref{tab:weakly_conditioned_urmp}). 
 
\begin{figure}[ht]
    \centering
    \includegraphics[width=\linewidth]{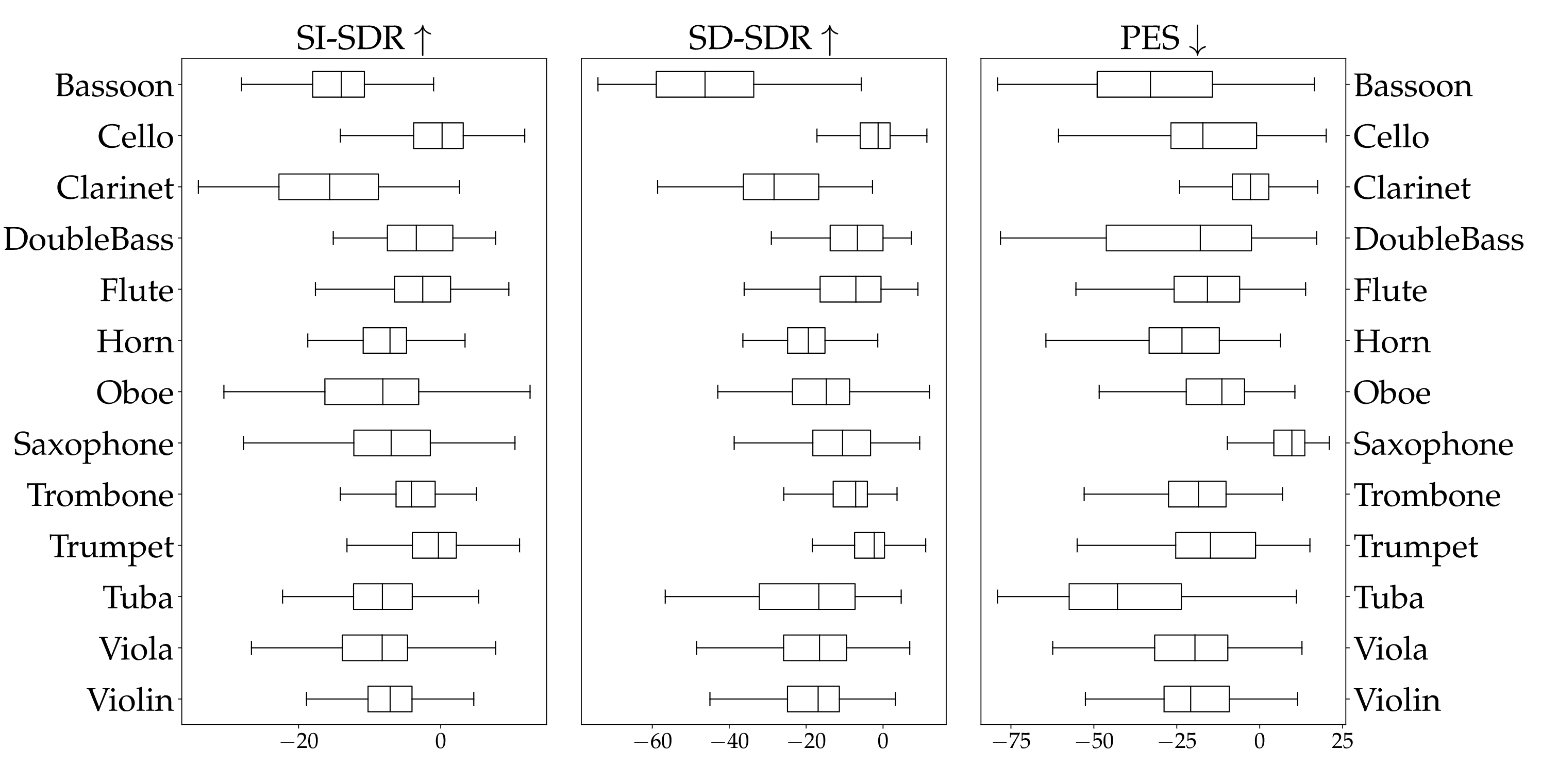}
    \caption{Exp. 11 per-instrument boxplots for the URMP dataset. Note that the x axis scale limits vary from metric to metric. The principal reason of the difference between SI-SDR and SD-SDR is that SD-SDR accounts for the volume changes.}
    \label{fig:urmp_per_instrument}
\end{figure}

Figure \ref{fig:urmp_per_instrument} shows the performance measured by SI-SDR, SD-SDR and PES in Exp. 11 for each instrument in the URMP dataset. The results emphasize the fluctuations between the instruments. As it can be observed, the worst SI-SDR and SD-SDR values correspond to clarinet, while the best three instruments in terms of SI-SDR are cello, trumpet, flute and double bass. Tuba is also the best instrument with respect to PES.
 
\begin{figure}[ht]
    \centering
    \includegraphics[width=\linewidth]{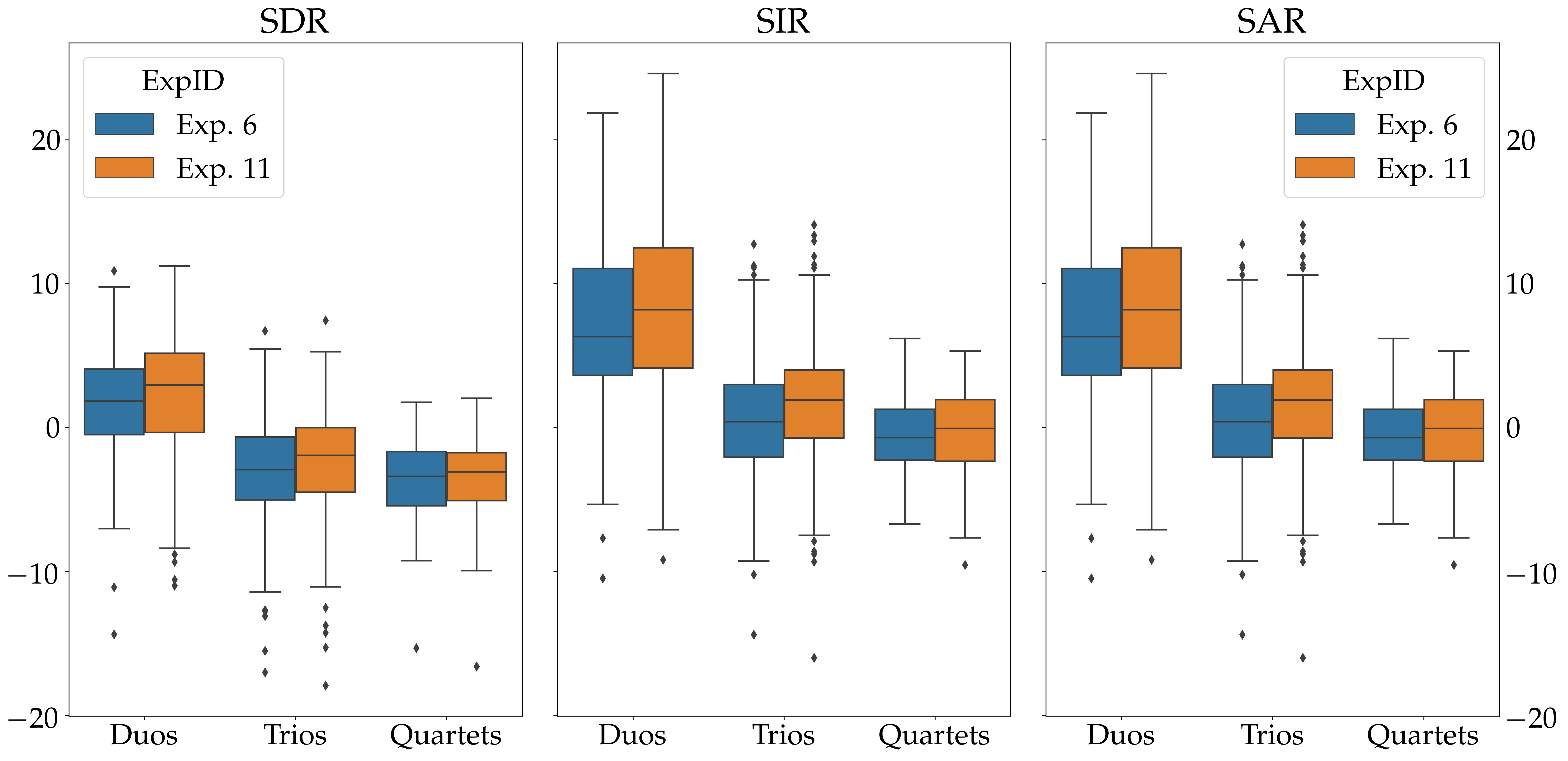}
    \caption{Exp. 6 (no conditioning) vs. Exp. 11 (weak label-conditioning, multiply) SDR, SIR and SAR boxplots for the URMP dataset with respect to the number of sources in the mixture. }
    \label{fig:urmp_per_n_source}
\end{figure}

On the other hand, Figure  \ref{fig:urmp_per_n_source} presents the SDR, SIR, and SAR metrics with respect to the number of sources in the mixture for two different models: the baseline network without conditioning (Exp. 6) and the same network with Label-multiply conditioning (Exp. 11). 
The improvements are consistent among the metrics and the number of voices in the mixture.


\begin{table*}[ht]
    \centering
\begin{tabular}{c||c|c|c|c|c|c|c|c}
        Method & Exp. ID & W & SI-SDR $\uparrow$ & SD-SDR $\uparrow$ & PES $\downarrow$ & SDR $\uparrow$ & SIR $\uparrow$ & SAR $\uparrow$ \\ \hline 
        w/o conditioning & 6 & 2 & $ -7.4\pm 9.9 $ & $ -17.5\pm 19.8 $ & $ \mathbf{ -24.4\pm 15.8} $ & $ -2.3\pm6.9 $ & $ 1.5\pm8.4 $ & $ 5.7\pm5.2 $ \\ 
        FiLM-bottleneck & 8 & 2 & $ -5.1\pm6.8 $ & $ -19.2\pm10.7 $ & $ -7.1\pm9.9 $ & \dotuline{$ -1.3\pm6.0 $} & $ 1.9\pm8.2 $ & $ 6.6\pm2.9 $ \\
        FiLM-encoder & 9 & 2 & $ \mathbf{-4.2\pm6.4} $ & $ -19.7\pm8.8 $ & $ -6.9\pm8.8 $ & $ \mathbf{-1.2\pm5.8} $ & \dotuline{$1.1\pm7.5$} & $\mathbf{8.5\pm3.0} $ \\
        FiLM-final & 10 & 2 & $ -4.3\pm6.3 $ & $ -17.3\pm9.1 $ & $ -9.3\pm11.2 $ & $-1.6\pm5.6$ & $ 0.6\pm7.3 $ & $ 8.2\pm3.0 $ \\
        Label-multiply & 11 & 2 & \dotuline{$ -7.2\pm9.8 $} & \dotuline{$ \mathbf{-15.3\pm15.5 }$} & $ -18.6\pm20.3 $ & $ -1.9\pm6.8 $ & $ \mathbf{2.4\pm8.8} $ & $ 5.1\pm4.1 $ \\

\end{tabular}
    \caption{Conditioned U-Net with Labels (URMP metrics). The best results are highlighted in {\bf bold}. Results that are {\bf not statistically significant} w.r.t. Exp.~6 ($ p < 0.05 $) are \dotuline{dotted}.}
    \label{tab:weakly_conditioned_urmp}
\end{table*}

\subsection{Conditioning on visual information}\label{ss:results:visual}

We compare visually conditioned U-Net with its corresponding non-conditioned and label conditioned baselines. 

Table \ref{tab:visual_conditioned_urmp} shows the performance of single-frame visually conditioned U-Net given the same FiLM locations as in the label conditioning case. It also indicates the results of conditioning by a visual-motion context vector learned from 5 frames per segment (with the frame rate set to 5/6 fps). Lastly, we report the results for the Sound-of-Pixels (SoP)\cite{zhao2018soundofpixels} method. SoP-unet7 states for the original method trained on the Music dataset published in \cite{zhao2018soundofpixels}. We used the officially provided weights and evaluated the model on the URMP dataset. SoP-unet7-ft indicates the version which was fine-tuned on the Solos dataset. SoP-unet5-Solos accounts for a model with 5 blocks in U-Net which is trained from scratch. In all SoP networks, both visual and audio networks are trained simultaneously while in our conditioning experiments \textcolor{black}{visual network is fine-tuned in all experiments with a single-frame conditioning and is frozen in all experiments with a multi-frame conditioning}. 

The metrics show that the best results both in terms of SI-SDR and SDR are attained by multiplicative conditioning at the final layer with a context vector extracted from a single visual frame (Exp. 16). Similar results are obtained using conditioning with visual-motion context vector (Exp. 13). The highest SAR among all visually conditioned U-Nets is achieved by a network FiLM-ed at the final layer (Exp. 18), while the latest also delivers  comparable results in SI-SDR,  SDR, and SIR to the best performing Final-multiply (Exp. 16).
Surprisingly, in case of visual conditioning, none of the models outperforms the baseline network without conditioning in terms of SD-SDR, PES, and SIR. Finally, compared to the SoP model (including its  fine-tuned version on Solos), all models with visual conditioning and the baseline without conditioning give better SI-SDR values. Most of those models outperform SoP as well in terms of SDR and SIR.


\begin{table*}[ht]
    \centering
    \begin{tabular}{c||c|c|c|c|c|c|c|c|c}
        Method & Frames & Exp. ID & W & SI-SDR $\uparrow$ & SD-SDR $\uparrow$ & PES $\downarrow$ & SDR $\uparrow$ & SIR $\uparrow$ & SAR $\uparrow$ \\ \hline 
        w/o conditioning & 0 & 6 & 2 & $ -7.4\pm 9.9 $ & $ \mathbf{-17.5\pm 19.8} $ & $ \mathbf{-24.4\pm 15.8} $ & $ -2.3\pm6.9 $ & $ \mathbf{1.5\pm8.4} $ & $ 5.7\pm5.2 $ \\
        FiLM-encoder & 1 & 12 & 2 & $ -12.1\pm13.9 $ & $ -34.9\pm29.7 $ & $ -21.7\pm22.3 $ & $ -5.1\pm7.9 $ & $ -0.3\pm8.1 $ & $ 3.3\pm6.9 $ \\

        FiLM-bottleneck & 1 & 17 & 2 & $ -5.1\pm7.7 $ & $ -17.6\pm14.7 $ & $ -16.3\pm17.7 $ & \dotuline{$-2.2\pm5.7$} & $ 0.2\pm7.0 $ & $ 7.2\pm3.7 $ \\
        
        FiLM-final & 1 & 18 & 2 & $ -4.8\pm7.3 $ & $ -23.7\pm20.2 $ & $ -21.3\pm18.7 $ & $-1.6\pm5.6$ & $ 0.8\pm8.0 $ & $ \mathbf{10.3\pm5.4} $ \\
        
        Final-multiply & 1 & 16 & 2 & $\mathbf{-4.1\pm7.2}$ & $ -19.0\pm13.9 $ & $ -15.4\pm14.3 $ & $ \mathbf{-1.5\pm6.1} $ & $ 1.0\pm8.4 $ & $ 9.2\pm3.7 $ \\ 
        
        \hdashline
        
        FiLM-final-lstm & 5 & 13 & 2 & $ -4.5\pm6.2 $ & $ -21.5\pm11.9 $ & $ -12.7\pm13.8 $ & \dotuline{$-2.1\pm5.4$} & $ -0.4\pm6.7 $ & $ 9.3\pm3.7 $ \\

        FiLM-bottleneck-lstm & 5 & 14 & 2 & $ -5.6\pm7.6 $ & $ -19.3\pm15.8 $ & $ -18.3\pm16.2 $ & \dotuline{ $-2.1\pm5.8$ } & $ 0.2\pm7.0 $ & $ 7.5\pm4.1 $ \\
        
        FiLM-bottleneck-mp & 5 & 15 & 2 & $ -8.2\pm9.0 $ & $ -26.6\pm21.4 $ & $ -19.3\pm17.7 $ & $ -3.3\pm6.3 $ & $ 0.5\pm7.6 $ & $ 5.2\pm5.5 $ \\ \hline

        SoP-unet7 \cite{zhao2018soundofpixels} & 3 & 19 & 0 & $-18.7\pm8.9$ & $-21.1\pm9.4$ & n/a & $-3.8\pm4.0$ & $-1.5\pm4.7$ & $7.6\pm3.1$ \\
        SoP-unet7-ft \cite{zhao2018soundofpixels} & 3 & 20 & 0 & $-17.5\pm8.5$ & $-20.3\pm9.3$ & n/a & $-2.6\pm5.0$ & $0.5\pm6.4$ & $6.9\pm2.5$ \\
        SoP-unet5-Solos & 3 & 21 & 0 & $-16.9\pm8.6$ & $-18.7\pm8.9$ & n/a & $-2.9\pm4.7$ & $-1.7\pm5.3$ & $11.1\pm6.9$ \\
    \end{tabular}
    \caption{Results for Visually Conditioned U-Net experiments with different types of conditioning and different number of frames used (evaluated on the URMP dataset). The best results are highlighted in {\bf bold}. Results that are {\bf not statistically significant} w.r.t. Exp. 6 ($ p < 0.05 $) are \dotuline{dotted}. }
    \label{tab:visual_conditioned_urmp}
\end{table*}

\subsection{Unsuccessful attempts}

We would like to report several strategies which did not improve source separation performance in our experiments. 

In one of the experiments, we used $L_2$ loss while directly predicting spectrogram values instead of using the masking-based approach. However, the network failed to converge. We hypothesise that this behaviour accounts for the higher complexity of the spectrograms and the sparsity of the outputs. Another potential cause might be the gradient\'s magnitudes that are not bounded as in the masking-based approach, therefore affecting the training dynamics of the network. The potential causes could be investigated in the future.

We also had an unsuccessful attempt to employ multi-task learning in order to further regularize the embedding space. In these experiments we jointly optimised classification and separation losses trying to predict which instruments are present in the mixture using the bottleneck U-Net features as an input for a small classifier consisting of a single fully-connected layer. While generally converging, the classification and separation performance were lower than the results of stand-alone models.

\subsection{Discussion}
\label{sec:discussion}

In our experiments we observe that the use of additional information (apart from the audio signal) can be beneficial for the separation performance. However, integrating domain knowledge might be challenging and has to be done carefully. In this work we try to go beyond the classical ``divide-and-conquer" approach in which data representations are learned separately and aggregated at the final stage of the separation pipeline. Nevertheless, our experiments show that most notable performance improvement can be achieved by multiplying learned masks by ground truth instrument labels or probabilities taken from visual data. Merging a strong label signal into the encoder seems to be helpful but also leads to overfitting.
From the results we observe that U-Net conditioned on the visual context vector performs worse in terms of SIR than the unconditioned version.  A possible explanation for this observation could be due to the confusions separating musical instruments from the same family (like viola and violin which have a strongly similar visual appearance). Those confusions may result in more interferences and mispredictions when both instruments are present in the mixtures, which is a common case for the URMP dataset. Indeed, as shown in Table \ref{tab:datasets}, violin and viola appear in great part of the duos, trios, and quartets of URMP dataset.

By inspecting the results obtained by the Sound-of-Pixels method, we highlight the importance of taking the source separation problem in the real-world scenario, as the method was previously tested in mix-and-separate settings and the reported results had an average SDR of 8dB. Our results demonstrate the demand for the testing on the real mixtures rather than using the mix-and-separate approach. Notably, even 5-blocks Sound-of-Pixels trained on Solos performs better than 7-blocks Sound-of-Pixels trained on Music. Fine-tuning of the original Sound-of-Pixels model allows to improve the quality of source separation for 1.2dB in SDR which also indicates the need of enlarging the datasets and enhancing their quality. 

Even though from the literature we know that source separation can benefit from integrating the motion information \cite{zhao2019soundofmotions, parekh2018guiding, li2017see}, we would like to note that all aforementioned methods use complex pre-processing in order to extract reliable motion features, which brings attention to the problem of closing the gap between motion and audio representations. 
Another fact that should be noted is that all sources of information should be correctly combined, preserving synchrony between them. While for single-frame visual and weak label information it is not so important, for temporal data such as motion, pitch, and musical scores it may become a crucial aspect for successful conditioning. Consecutively, a different baseline source separation architecture, such as an RNN-based network, may improve on the current results due to its sequential nature which better preserves time-domain information.

Taking into consideration the above-mentioned observation, we can note that the U-Net architecture may be a limitation of our study, and the results may be different for other baseline architectures. 

Given that the best results in terms of different metrics are achieved by using different setups (e.g. binary and ratio masks), we would like to emphasise that a further enhancement can be obtained by having the best of both worlds. This would require an additional network that combines the estimates
of binary and soft masks, as proposed in \cite{grais2016combining}. 
Finally, we would like to note the opportunity to surpass the current performance by employing additional constraints for the loss functions as in \cite{gao2019co}, \cite{wisdom2019differentiable}, or weighting the loss values of the masks with the magnitude values of the mixture \cite{zhao2018soundofpixels, gao2019co} as it may help to avoid treating every time frequency bin equally and focus attention on the areas where most of the energy is concentrated.

\section{Conclusion}\label{ss:conclusion}

We tackle a problem of Single Channel Source Separation for multi-instrument polyphonic music conditioned on external data. In this work we have shown that the use of extra information such as (1) binary vectors indicating the presence or absence of musical instruments in the mix and (2) visual feature vectors extracted from corresponding video frames improve the separation performance.

We also show that different types of conditioning have different effects w.r.t. the performance metrics. We have conducted a thorough study of FiLM-conditioning introduced at three possible locations of the primary source separation U-Net model. We have demonstrated that the best results in SI-SDR and SDR can be obtained with label conditioning by FiLM at the encoder or by visual conditioning at the final output mask.

The results shown in the present work indicate that the real-case scenario such as chamber quartets source separation is challenging and there is still a significant performance gap of about 13dB between the state-of-the-art separation methods and ideal ratio masks.

Potential improvements could include modifying the U-Net architecture, combining binary and soft masks to obtain a good balance between SDR and PES. Another possibility could be integrating an advanced motion analysis network and employing audio-motion synchrony for conditioning the network, and conditioning on musical scores.

\appendices
\section{Hyperparameters of the experiments} \label{app:hyperparams}

We provide the full set of model hyperparameters used in the experiments in Section  \ref{ss:results} in Table \ref{tab:cunet_ablation_parameters}. 
Please note that only one parameter changes between each pair of the experiments compared in Table \ref{tab:baseline_urmp}. For the experiments in Section \ref{ss:results:visual} the model parameters are set as described in Section \ref{ss:exp_ablation}.

\begin{table*}[ht]
\centering
\begin{tabular}{c ||c|c|c|c|c|c|c|c}
Exp. ID & STFT F-scale & STFT V-scale & model & noise & mask & loss & CL & conditioning  \\ \hline 

1 & log & dB-norm & U-Net & No & Ratio & $L_2$ & No & None \\ 
2 & log & dB-norm & U-Net & No & Ratio & $L_2$ & Yes & None \\ 
3 & log & dB-norm & U-Net & No & Binary & BCE & No & None\\ 
4 & log & dB-norm & U-Net & Yes & Ratio & $L_2$ & No & None \\ 
5 & log & log & U-Net & No & Ratio & $L_2$ & No & None \\ 
6 & linear & dB-norm & U-Net & No & Ratio & $L_2$ & No & None \\ 
7 & log & dB-norm & MHU-Net & No & Ratio & $L_2$ & No & None \\         \hdashline

8 & linear & dB-norm & U-Net & No & Ratio & $L_2$ & No & FiLM-bottleneck (binary)\\ 

9 & linear & dB-norm & U-Net & No & Ratio & $L_2$ & No & FiLM-encoder (binary) \\ 
10 &  linear & dB-norm & U-Net & No & Ratio & $L_2$ & No & FiLM-final (binary) \\ 
11 & linear & dB-norm & U-Net & No & Ratio & $L_2$ & No & Label-multiply (binary) \\ \hdashline
12 & linear & dB-norm & U-Net & No & Ratio & $L_2$ & No & FiLM-encoder (visual) \\ 
13 & linear & dB-norm & U-Net & No & Ratio & $L_2$ & No & FiLM-final-lstm (visual-motion) \\ 
14 & linear & dB-norm & U-Net & No & Ratio & $L_2$ & No & FiLM-bottleneck-lstm (visual-motion)  \\ 
15 & linear & dB-norm & U-Net & No & Ratio & $L_2$ & No & FiLM-bottleneck-maxpool (visual-motion) \\ 
16 & linear & dB-norm & U-Net & No & Ratio & $L_2$ & No & Final-multiply (visual) \\ 
17 & linear & dB-norm & U-Net & No & Ratio & $L_2$ & No & FiLM-bottleneck (visual) \\ 
18 & linear & dB-norm & U-Net & No & Ratio & $L_2$ & No & FiLM-final (visual) \\ \hdashline
19 & log & log & SoP-7 & No & Binary & BCE & No & SoP-LinearCombination (visual) \\ 
20 & log & log & SoP-7-ft & No & Binary & BCE & No & SoP-LinearCombination (visual) \\ 
21 & log & log & SoP-5 & No & Binary & BCE & No & SoP-LinearCombination (visual) \\ 

\end{tabular}
    \caption{Hyperparameters for experiments reported in this study and corresponding experiment IDs for Conditioned U-Net. CL denotes the usage of curriculum learning.}
    \label{tab:cunet_ablation_parameters}
\end{table*}

\section{Per-experiment bar plots with source separation performance results}\label{app:urmp_barplots}

\begin{figure*}[ht]
    \centering
    \includegraphics[width=\linewidth]{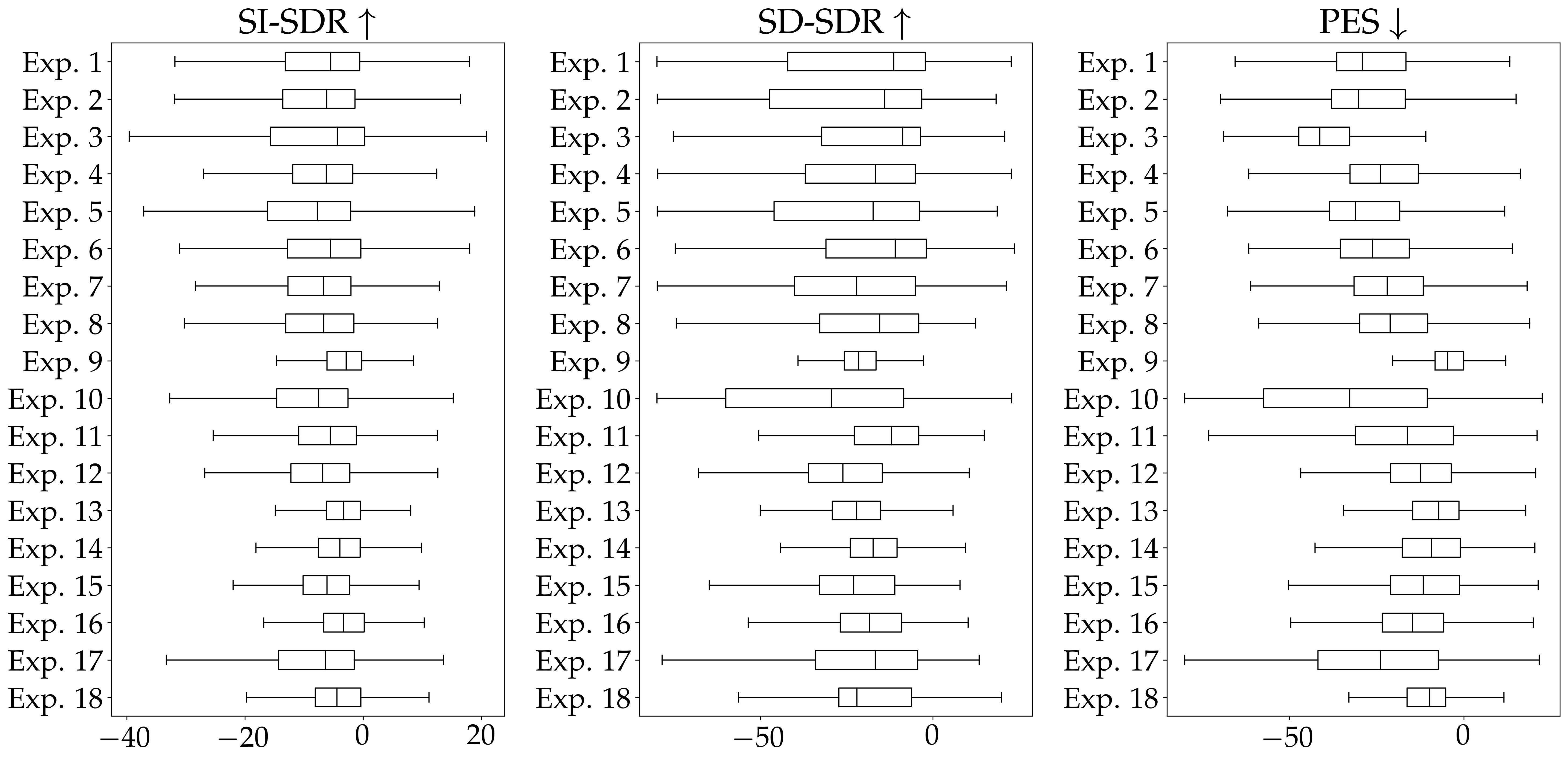}
    \caption{SI-SDR, SD-SDR and PES boxplots for the experiments from Section \ref{ss:results}. Experiments are referenced by ID. }
    \label{fig:urmp_audio_only}
\end{figure*}

Figure \ref{fig:urmp_audio_only} shows source separation metrics (SI-SDR, SD-SDR, PES) pictured as bar plots with mean and standard deviation for each experiment conducted in Section \ref{ss:results}. The experiment are referenced by id as in Tables~\ref{tab:baseline_urmp}, \ref{tab:weakly_conditioned_urmp} and \ref{tab:visual_conditioned_urmp}.

\section*{Acknowledgment}

This work was funded in part by ERC Innovation Programme (grant 770376, TROMPA); Spanish Ministry of Economy and Competitiveness under the Mar{\'i}a de Maeztu Units of Excellence Program (MDM-2015-0502) and the Social European Funds; the  MICINN/FEDER UE project with reference PGC2018-098625-B-I00; and the H2020-MSCA-RISE-2017 project with reference 777826 NoMADS. 
We gratefully acknowledge NVIDIA for the donation of GPUs used for the experiments.

\newpage

\bibliographystyle{IEEEtran}
\bibliography{bibliography}

\newpage

\begin{IEEEbiographynophoto}{Olga Slizovskaia} received the M.Sc. degree in computational mathematics and cybernetics from Moscow State University, Moscow, Russia, in 2013. She has been working towards the Ph.D. degree in audio-visual music information retrieval at Universitat Pompeu Fabra (UPF), Barcelona, Spain, since 2016. Her research interests include developing machine learning and deep learning methods for audio and multimodal data analysis.
\end{IEEEbiographynophoto}
\begin{IEEEbiographynophoto}{Gloria Haro} is an associate professor at Universitat Pompeu Fabra (UPF, Spain). She finished her Ph.D. at UPF in 2005. She was a postdoctoral fellow at the Institute for Mathematics and its Applications, University of Minnesota (USA, 2005-2007), a "Juan de la Cierva" fellow at Universitat Politècnica de Catalunya (Spain, 2007-2008), and a "Ram{\'o}n y Cajal" fellow at UPF (Spain, 2008-2014). Her current research is focused on video understanding and automatic video editing tools for post-production, which involves research in basic problems such as optical flow estimation, video inpainting, segmentation, and depth estimation. She is also working in audio-visual approaches for video understanding, such as instrument recognition, source separation and sound localisation.
\end{IEEEbiographynophoto}
\begin{IEEEbiographynophoto}{Emilia G{\'o}mez} (Telecommunication Engineer, PhD in Computer Science) is an associate professor at Universitat Pompeu Fabra in Barcelona. She also leads the Human Behaviour and Machine Intelligence (HUMAINT) project at the Centre for Advanced Studies of the Joint Research Centre, European Commission. During her career, Emilia G{\'o}mez has developed data-driven machine learning algorithms to support music listening experiences. Starting from music, she researches on the social, cultural and ethical impact of Artificial Intelligence systems and related aspects such as diversity, fairness and trust.\end{IEEEbiographynophoto}

\end{document}